\documentclass[aps,prb,twocolumn,preprintnumbers,amsmath,amssymb,10pt,superscriptaddress]{revtex4-1}
\usepackage[colorlinks=true,linkcolor=blue,anchorcolor=red,citecolor=blue, urlcolor=blue]{hyperref}
\usepackage{graphicx}
\usepackage{color}
\usepackage{multirow}
\usepackage{extarrows}
\usepackage{cases}

\def\avg#1{\left\langle#1\right\rangle}

\def\abs#1{\left|#1\right|}

\def\be{\begin{equation}}       \def\ee{\end{equation}}
\def\bea{\begin{eqnarray}}      \def\eea{\end{eqnarray}}
\def\ba{\begin{array}}
\def\ea{\end{array}}
\def\bnum{\begin{enumerate} }
\def\enum{\end{enumerate}}

\def\nn{\nonumber}

\def\=>{\Rightarrow}
\def\>{\rightarrow}

\def\eye2{Fathbb{I}}

\def\m{\textrm{matter}}

\def\eq#1{Eq.~(\ref{#1})}

\renewcommand{\>}{\rangle}

\newcommand{\GL}{{\rm GL}(1|1)}

\begin{document}
\title{Global phase diagram of the one-dimensional Sachdev-Ye-Kitaev model at finite $N$}
\author{Xin Dai}
\affiliation{Institute for Advanced Study, Tsinghua University, Beijing 100084, China}
\affiliation{Department of Physics, Ohio State University, Columbus, Ohio 43210, USA}
\author{Shao-Kai Jian}
\affiliation{Institute for Advanced Study, Tsinghua University, Beijing 100084, China}
\affiliation{Condensed Matter Theory Center, Department of Physics, University of Maryland, College Park, Maryland 20742, USA}
\author{Hong Yao}
\email{yaohong@tsinghua.edu.cn}
\affiliation{Institute for Advanced Study, Tsinghua University, Beijing 100084, China}
\affiliation{State Key Laboratory of Low Dimensional Quantum Physics, Tsinghua University, Beijing 100084, China}

\date{\today}

\begin{abstract}
Many key features of the higher-dimensional Sachdev-Ye-Kitaev (SYK) model at {\it finite} $N$ remain unknown. Here we study the SYK chain consisting of $N$ ($N$$\ge$$2$) fermions per site with random interactions and hoppings between neighboring sites. In the limit of vanishing SYK interactions, from both supersymmetric field theory analysis and numerical calculations we find that the random hopping model exhibits Anderson localization at finite $N$, irrespective of the parity of $N$. Moreover, the localization length scales linearly with N, implying no Anderson localization \textit{only} at $N\!=\!\infty$. For finite SYK interaction $J$ , from the exact diagonalization we show that there is a dynamic phase transition between many-body localization and thermal diffusion as $J$ exceeds a critical value $J_c$. In addition, we find that the critical value $J_c$ decreases with the increase of $N$, qualitatively consistent with the analytical result of $J_c/t \!\propto\! \frac{1}{N^{5/2}\log N}$ derived from the weakly interacting limit.
\end{abstract}
\maketitle

\section{Introduction}
The seminal Sachdev-Ye-Kitaev (SYK) model~\cite{sachdev93prl,kitaev2018soft} presents a zero-dimensional cluster consisting of $N$ Majorana fermions with random all-to-all interactions. In the large-$N$ limit it is exactly solvable, exhibiting maximal quantum chaos \cite{kitaev2018soft,Maldacena16prd,maldacena2016bound}, emergent $SL(2,R)$ symmetry as well as a holographic dual to dilaton gravity theory in nearly AdS$_2$ geometry~\cite{kitaev2018soft,Maldacena16prd}. Owing to its solvability and intriguing properties, it has stimulated enormous excitement \cite{
polchinski2016spectrum,  fu2016numerical,  bi2017instability, Altman17prb, garcia2016spectral, cotler2017black, garcia2017analytical, chen2017tunable,  bagrets2017power, khveshchenko2017thickening, eberlein2017quantum, sonner2017eigenstate, gu2017spread, liu2017quantum,  huang2017eigenstate, kitaev2018soft, jia2018exact, garcia2018many, tarnopolsky2018large, Maldacena16progress, Jensen16prl, Verlinde16jhep,gross17jhep, jevicki2016bi, das18jhep,Das17jhep,cai17arxiv,jian17prb,Blake2018,  gu17jhep, Davison17prb, lucas17scipost, Song17prl, berkooz2017higher, Jian17prl, CMJian17prb, chen17prl,zhang2017dispersive, Cai2018,garcia2017,patel17arxiv,chowdhury2018translationally,wu2018candidate}. In particular, the large-$N$ limit of the SYK model, after properly generalized to higher dimensions \cite{gu17jhep,Davison17prb, lucas17scipost,Song17prl,berkooz2017higher,Jian17prl, CMJian17prb, chen17prl,zhang2017dispersive, Cai2018, garcia2017,patel17arxiv, khveshchenko2017thickening, chowdhury2018translationally, wu2018candidate}, could provide an insightful and promising avenue to investigate the spectral and transport properties of non-Fermi liquid states. Nonetheless, features of the higher-dimensional SYK models with {\it finite} $N$ remain largely unknown. As the case of finite $N$ is directly relevant to possible experimental realizations \cite{ippei2017, Franz, jason2017, xu2017} of SYK models, it is desired to understand the characterizing properties of the higher-dimensional SYK models at finite $N$.

\begin{figure}[t]
\includegraphics[width=4.1cm]{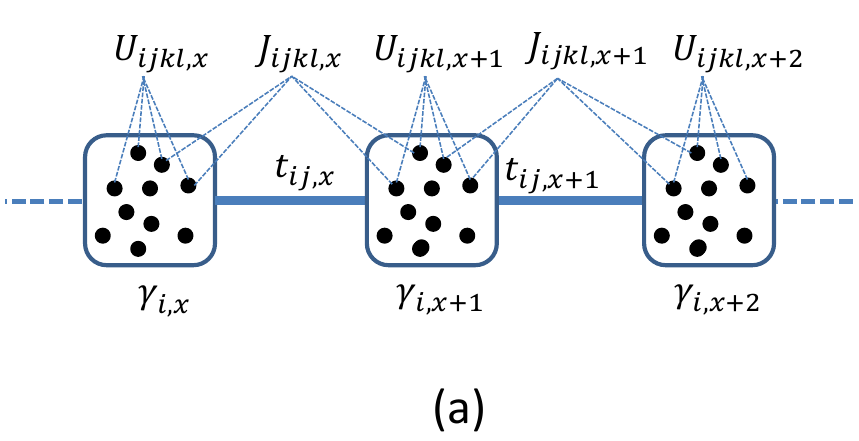}~~
\includegraphics[width=4.cm]{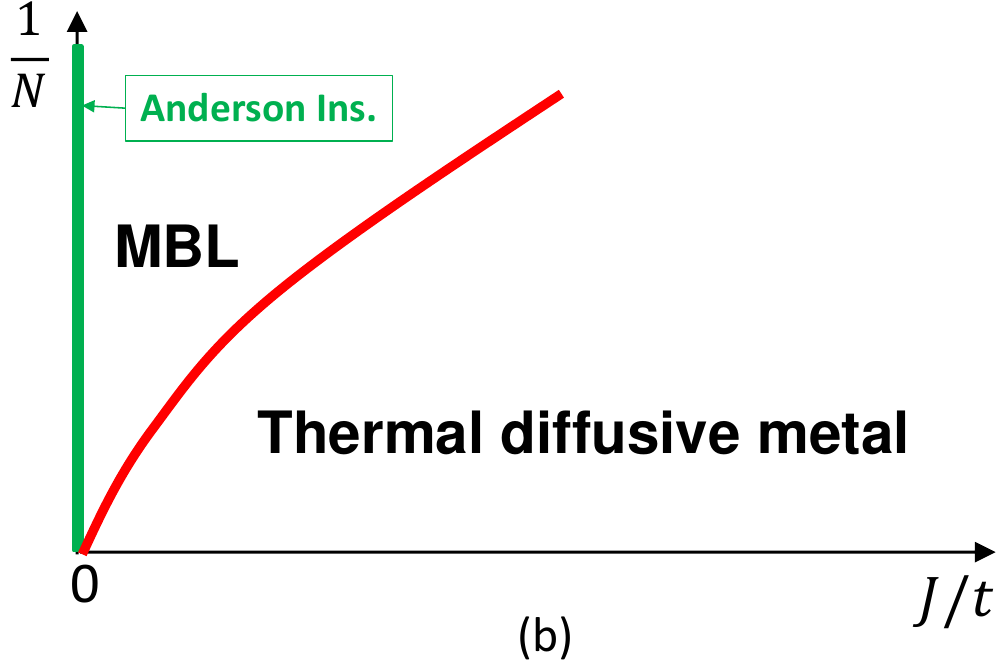}
\caption{\label{fig1} (a) The schematic representation of the SYK chain model. (b) The global phase diagram of the SYK chain model at finite $N$.  The system is in the many-body localized (MBL) phase when the SYK interaction $J$ is relatively weak, but exhibits thermalization and diffusion when $J$ exceeds a critical value $J_c$. Note that the critical value $J_c$ decreases with increasing $N$.}
\end{figure}

Here we consider a generic SYK chain model of Majorana fermions respecting time-reversal symmetry, which includes four-fermion random interactions and random hoppings between neighboring sites as shown in Fig. \ref{fig1} [see \eq{realSYK} below]. Note that the neighboring fermion hopping on a bipartite lattice respects the time-reversal symmetry defined as $\gamma_{j,x} \rightarrow (-)^x \gamma_{j,x}$ where $\gamma_{j,x}$ represents the Majorana fermion with flavor $j\!=\!1,\cdots\!,N$ on site $x$.  Both the random hoppings and the random interactions are characterized by Gaussian random variables with zero mean; their variances are given by $t^2/N$ and $3!J^2/N^3$, respectively.

We first consider the noninteracting limit, namely $J\!=\!0$, for which the model in \eq{realSYK} reduces to a one-dimensional (1D) random hopping model~\cite{Altland01nulear}. The presence of time-reversal symmetry renders the Majorana system in the BDI class~\cite{kitaev2009, ryu2010}.
In particular, when the system size $L$ is odd, there will be $N$ zero-energy single-particle modes in the band center due to the particle-hole symmetry. From numerical calculations, we find that the zero modes are localized for finite $N$ (both even and odd), which implies that all single-particle wavefunctions are Anderson localized. Moreover, our results show that the localization length scales linearly with the fermion flavor $N$, i.e., $\xi\!\propto\! N$, indicating the absence of Anderson localization {\it only} at $N\!=\!\infty$. Inspired by the pioneering work of Refs.~\cite{Altland99nuclearB, Atland99duplicate}, we further derive the corresponding supersymmetric field theory and find that the low energy physics can be described by the supersymmetric nonlinear $\sigma$ model with a {\it vanishing} topological $\theta$ term. From supersymmetric field-theory analysis, we obtain that the corresponding conductance decays exponentially with system size and the localization length scales linearly with $N$,
consistent with the numerical calculations.

For the case of finite interactions, by performing exact diagonalization (ED) we show that there is a dynamic phase transition from the many-body localized (MBL) phase~\cite{Gornyi05prl, basko06annal, huse07prb, nandkishore2015many} to the thermal diffusive metal phase as the interactions strength exceeds a critical value $J_c$. When $J<J_c$, the tendency to the MBL phase can be understood perturbatively:  a weak interaction is irrelevant to the Anderson localized phase in the noninteracting limit so the system remains many-body localized; namely sufficiently weak SYK interactions cannot effectively thermalize the system which is Anderson localized in the noninteracting limit. 
Numerically we find that the dynamic phase transition is characterized by the critical 
exponent $\nu\approx 1.1\pm 0.1$. which is consistent with previous works on MBL transition using small system size ED.
Moreover, as shown in Fig. \ref{fig1}b, we find that the critical interaction strength $J_c$ needed to thermalize the system decreases with the increase of $N$, which is consistent with the analytical result of $J_c/t \!\propto\! \frac{1}{N^{5/2}\log N}$ derived from the weakly interacting limit~\cite{basko06annal}.

\section{Model}
We consider the SYK chain model of Majorana fermions,
\bea\label{realSYK}
\hat H &=& \sum_{x,jk} i t_{jk,x} \gamma_{j,x} \gamma_{k,x+1}
 +\sum_{x,ijkl} J_{ijkl,x} \gamma_{i,x} \gamma_{j,x} \gamma_{k,x+1} \gamma_{l,x+1}\nn\\
&&~~+ \sum_{x,ijkl}U_{ijkl,x} \gamma_{i,x} \gamma_{j,x}\gamma_{k,x} \gamma_{l,x},
\eea
where $\gamma_{j,x}$ represent Majorana fermions with flavor index $j\!=\!1,\cdots,N$ on site $x\!=\!1,\cdots,L$. Here $U_{ijkl,x}$ label the usual on-site SYK interactions while $t_{jk,x}$ and $J_{ijkl,x}$ refer to random hopping and interaction between neighboring sites that are Gaussian random variables with mean $t_0=0$ and variance $\langle t_{jk,x}^2 \rangle \!=\! t^2/N$
and $\langle J_{ijkl,x}^2\rangle\!=\! J^2/N^3$, respectively.
The hopping of Eq.~\eqref{realSYK} represents the random hopping model in the strong disorder limit ($t/t_0\!=\!\infty$). We shall show below that it is qualitatively different from
the weak disorder limit ($t\!\ll\! t_0$) in terms of the localization physics studied in the literature~\cite{Altland01nulear,Brouwer98prl,Balents97prb,Brouwer2000prl}.
It is obvious that the model in \eq{realSYK} respects the time-reversal symmetry defined as $\gamma_{j,x}\!\to\! (-1)^j \gamma_{j,x}$. The time-reversal invariance then forbids onsite quadratic term $i\gamma_{i,x}\gamma_{j,x}$ in the Hamiltonian.

In the following, we shall focus on the case of vanishing onsite interactions, namely $U_{ijkl,x}\!=\!0$, while varying the nearest-neighbor SYK interaction strength $J$ with respect to the hopping strength $t$. This is partly because the onsite SYK interactions cannot be defined for the case of $N\!=\!2$ Majorana fermions. In contrast, a finite nearest-neighbor SYK interaction $J$ is allowed for all $N\!\ge\! 2$, including $N\!=\!2$. As the case of $N\!=\!2$ is numerically more accessible, we can obtain more reliable results up to a reasonably large system size $L$. Nonetheless, we would like to emphasize that the general feature of the global phase diagram and the universal properties of the MBL transitions do not depend on the specific SYK interactions we consider. In other words, we expect that characters of the phase diagram and the transitions obtained for the nearest-neighbor SYK interactions also apply to the case of onsite SYK interactions. As an illustration, we calculate the many-body level statistics with solely onsite SYK interactions for $N=4$, the result of which shown in Fig. S2 of the Supplemental Material is qualitatively the same as that of the nearest-neighbor SYK interactions. Consequently, we study the phase diagram as a function of $N$ and $J/t$ to include the case of $N\!=\!2$, while setting the onsite interaction to be zero.

\section{The noninteracting limit}
\begin{figure}[tbp]
\begin{minipage}{0.49\linewidth}
\centering
\includegraphics[width=0.97\columnwidth,height=3.05cm]{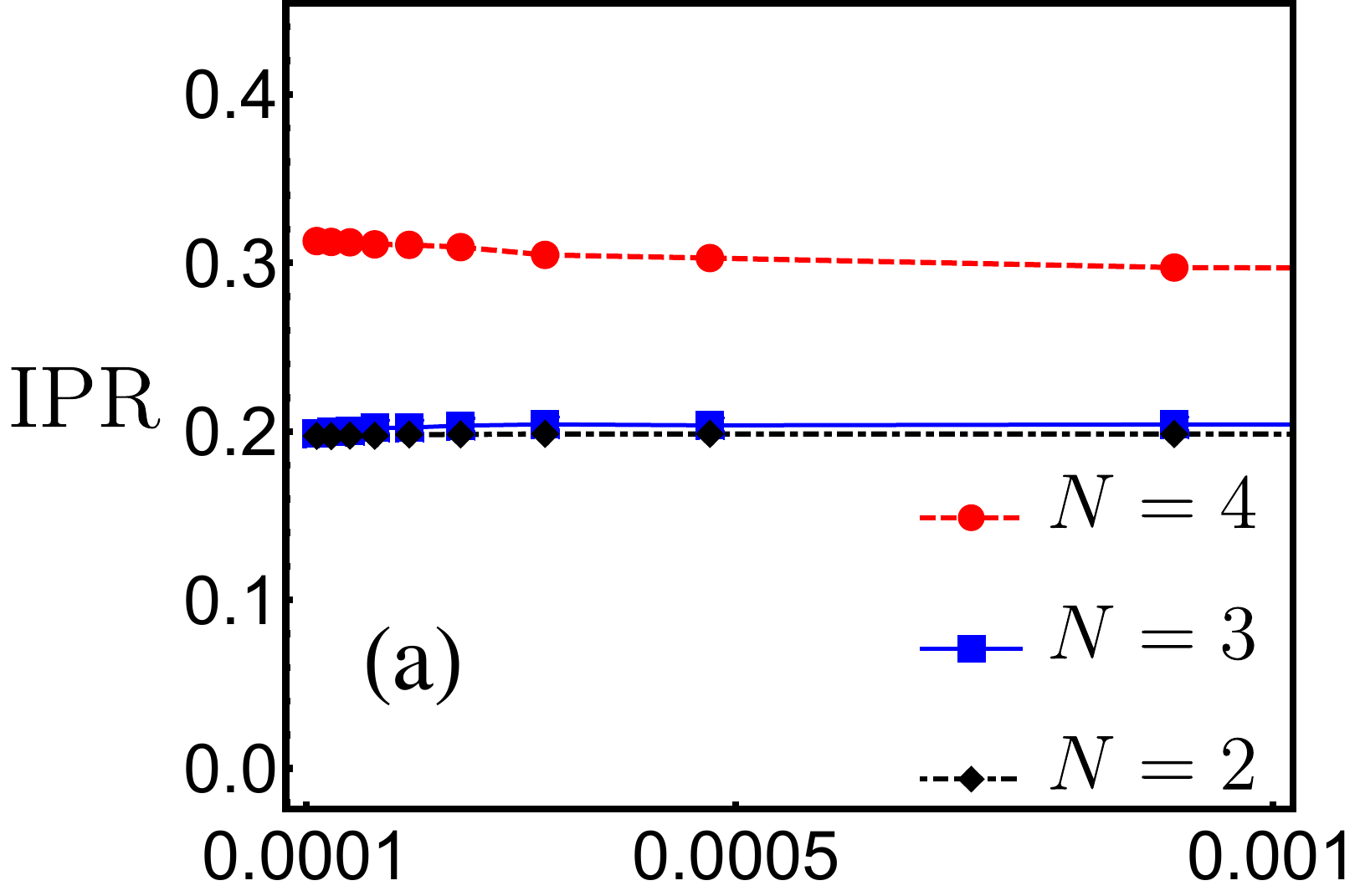}
\par \medskip \vfill
\includegraphics[width=0.95\columnwidth,height=3.15cm]{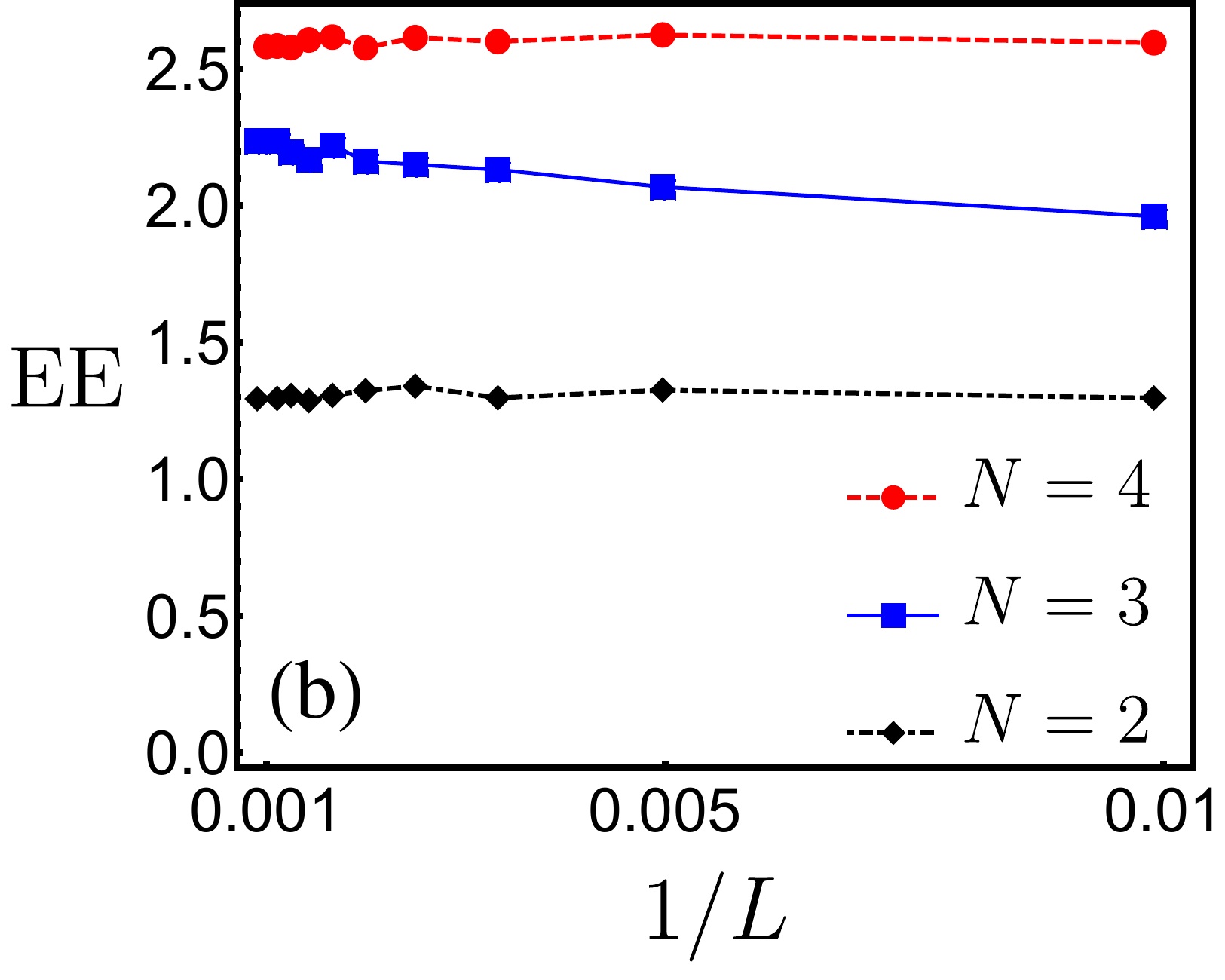}
\end{minipage}~
\begin{minipage}{0.49\linewidth}
\centering
\includegraphics[width=1.0\columnwidth,height=6.4cm]{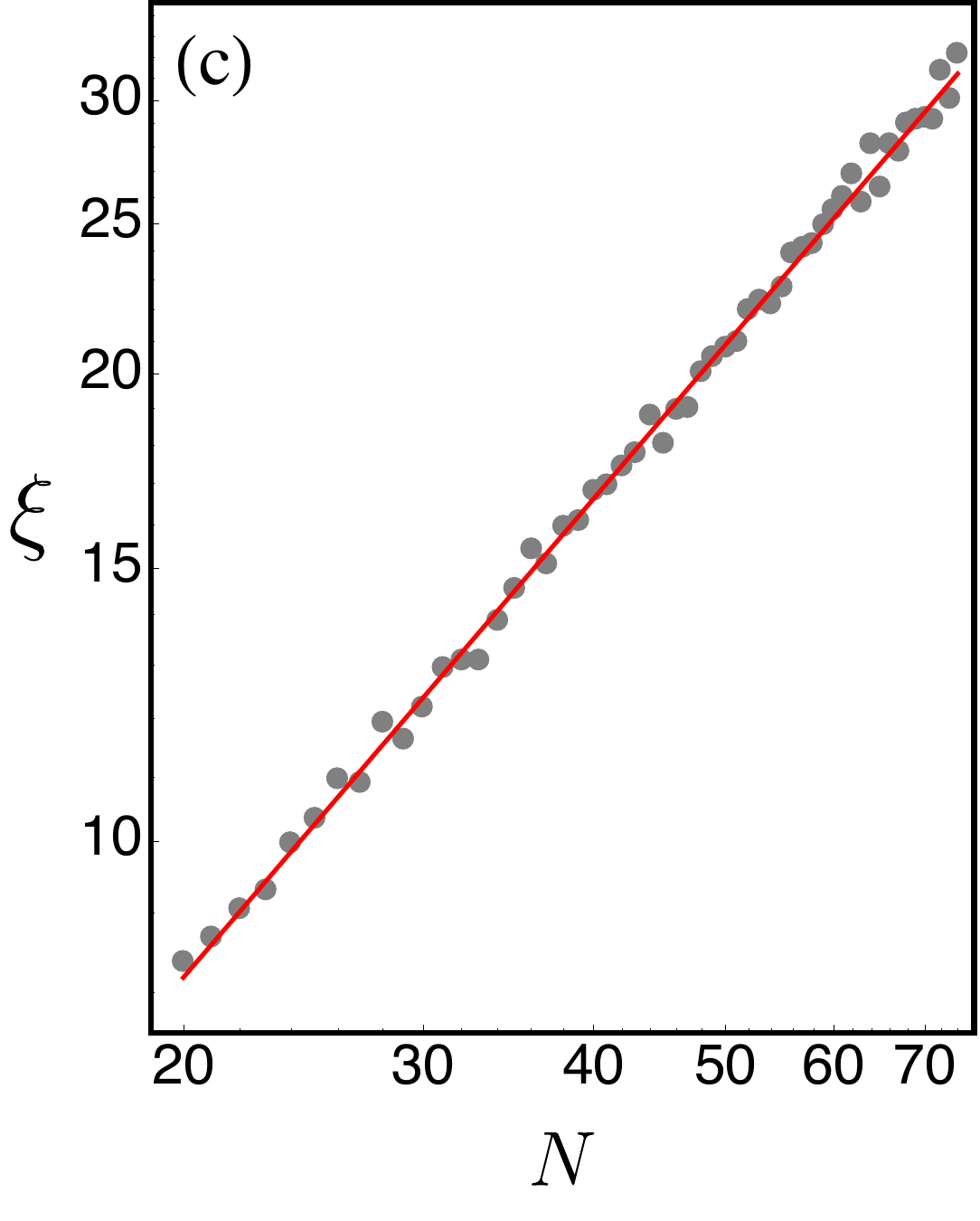}
\end{minipage}%
\caption{(a) For $N=2,3,4$, we compute both the scaling behavior of disorder averaged IPR (a)
and ground-state entanglement entropy (EE) (b) with system size $L$. The Fermi level is set to zero in computing the half-chain EE. (c) The representative linear fit of $\xi$ with $N$ for $N=20,21,\cdots,76$ after 600 disorder realizations with $L=6001$. For clarity, we only show the scaling behavior of one zero mode for each $N$ and the results for all other zero modes are similar.}
\label{xi-scaling}
\end{figure}

In the noninteracting limit, Eq. (\ref{realSYK}) is equivalent to the random hopping model in the strong disorder limit. It was shown previously that, when $N$ is odd, the zero modes of the random hopping model in the weak disorder limit are extended rather than Anderson localized \cite{Altland01nulear,Brouwer98prl,Balents97prb,Brouwer2000prl}, it is not known if the system is Anderson localized or not in the current strong disorder case,
especially for odd $N$ Majorana fermions. Thus, we numerically calculate the inverse participation ratio (IPR) \cite{Bhatt12prl} of the zero-mode wavefunction, which is defined by $\textrm{IPR}=
\frac{\sum_{x=1}^L(\psi^*_x \psi_x)^2}{\left(\sum_{x=1}^L\psi^*_x\psi_x\right)^2}$,
where $\psi_x$ labels a zero-mode wavefunction and $L$ denote the lattice size. Towards the thermodynamic limit $L\!\to\! \infty$, the scaling behaviors of the ensemble averaged IPR can tell if the wavefunction is localized ($\textrm{IPR}\propto\textrm{const.}$), extended ($\propto\!\frac{1}{L}$), or critical ($\propto\!\frac{1}{L^{\zeta}}$ with $0\!<\!\zeta\!<\!1$). As shown in Fig.~\ref{xi-scaling}(a), the IPR saturates to some nonzero constants with increasing $L$ for $N\!=\!2,3,4$, signaling a very strong localization behavior.

As a benchmark, we also study the scaling behaviors of the hal-chain entanglement entropy (EE) of the ground-state wavefunction of the random hopping chain using Klich's method~\cite{klich06}. It was shown in Refs.~\cite{Pouranvari13prb,Pouranvari14prb} that in the noninteracting system inspecting entanglement properties of the ground-state {\it alone} can tell if the system is localized or not. As shown in Fig.~\ref{xi-scaling}(b), the ground-state EE saturates to a constant value as $L\!\to\!\infty$ for $N\!=\!2,3,4$, implying a localized state.

The scaling behaviors of both the IPR and ground-state EE with respect to the system size $L$ yield consistent results and suggest that the single particle wavefunctions are Anderson localized in the noninteracting limit, in contrast to the case with constant diagonal hopping \cite{Altland01nulear}. To see if the Anderson localization persists to larger $N$,
we compute the IPR of the zero modes up to $N=76$ with fixed system size. The corresponding localization lengths can be extracted from the relation $\textrm{IPR}\propto 1/\xi$~\cite{Bhatt12prl}. From the log-log plot shown in Fig.~\ref{xi-scaling}(c), we find that the localization length $\xi$ of $N\in[20,76]$ can fit linearly with $N$, namely $\xi\! \propto\!  N$ for $N\gg 1$. It is quite remarkable that a single linear fit works for both even and odd $N$; no discernible sign of parity oscillations can be observed. Note that this linear scaling relation of localization length holds for all zero-mode wave functions.

Although a similar relation was observed in the weak disorder limit ($t\!\ll\! t_0$)~\cite{Altland01nulear, Brouwer98prl, Altland15prb},
there is an important and qualitative distinction with the present
strong disorder limit ($t/t_0\!=\!\infty$). For the case of the weak disorder limit, Anderson localization occurs only for even $N$ while all zero energy wavefunctions are extended for odd $N$. Consequently, it is natural to infer that the topological protection of the delocalization in the wavefunction for odd $N$ in the weak disorder limit fails in the strong disorder limit. Indeed, as we shall show below, the topological $\theta$ term in the supersymmetric nonlinear-$\sigma$ model vanishes in the strong-order case for both even- and odd $N$, consistent with the numerical results discussed above.

{\bf Supersymmetric field theory:} To furnish a firm understanding of numerical results, we develop a field theory using the supersymmetry approach~\cite{Efetov_book, Mirlin2000,Altland99review,wegner2016book} which is a powerful tool for analyzing noninteracting disorder problems. For simplicity, we only sketch the derivation and the details can be found in the SM. While the supersymmetry method was originally developed to deal with complex fermions, concerning the single particle physics the results of the supersymmetry theory apply for both complex and Majorana fermions as we argue below. Suppose the single particle Hamiltonian for Majorana fermions takes the form of $H(\gamma)= \sum it_{jk,x}\gamma_{j,x} \gamma_{k,x+1}$. Imagine there exists an identical ``ghost'' copy $H(\gamma')$ of the original $H(\gamma)$ such that they add up forming the complex fermionic Hamiltonian $H(\chi)\!=\!H(\gamma)+H(\gamma') \!=\!\sum_{jk} [it_{jk,x} \chi^\dag_{j,x}\chi_{k,x+1}+H.c.]$ where $\chi_j\!=\!(\gamma^1_j+i\gamma^2_j)/2$ are complex fermion annihilation operators. The localization properties of the complex fermion model $H_(\chi)$ are identical to those of the Majorana fermion model $H(\gamma)$ as they share the same single-particle matrix $it_{jk,x}$.

The basic idea of the supersymmetry method is to promote the original anticommuting fermionic field $\chi$ to the superfield $\psi$ by adding a commuting bosonic counterpart $\phi$, i.e., $\psi=(\phi,\chi)^T$, such that the disorder average can be performed at the very beginning, due
to the cancellation of determinants from the Gaussian integrals of complex and Grassmann variables. After the disorder average, the partition function can be written as
\bea\label{partition_function}
Z\!=\!\int\!{\cal D}(\bar\psi\!,\!\psi)
\exp\Big[i\sum_{n} \bar\psi_{n,\mu}z\psi_{n,\mu}
\!-\!\frac{2t^2}{N}\!\sum_{\substack{n\in{A}\\m\in{B}}}
{\rm str\;} g^{\mu\mu}_n g^{\nu\nu}_m\Big],~~~
\eea
where summation over repeated indices is assumed, $z$ is the frequency, ${\rm str}$ represents the supertrace, and $g^{\mu\mu}_n\equiv \psi_{n,\mu}\otimes\bar\psi_{n,\mu}$ is the superfield bilinear living on $A$ and $B$ sublattices, respectively (for details see the Supplemental Materials). To proceed, we introduce two auxiliary supermatrix fields $Q^{\pm}_{nm}\equiv Q_{A,n}\pm iQ_{B,m}$ to decouple the quartic term and then integrate out the superfield $\psi$ to obtain the action in terms of the superfield $Q$. The next step is to get the saddle point solution $\frac{\delta S}{\delta Q^{\pm}}=0$. Then we perform gradient expansions around the ground-state manifold to identify the low energy degrees of freedom. The resulting effective action at $z=0$ is
\begin{equation}
S[T] = -\frac{\tilde{\xi}}{8}\int dr {\rm str}( \partial T^{-1} \partial T ),
\label{effective_action}
\end{equation}
where $\tilde{\xi}=N$ is in units of the lattice constant $a$.

One key feature of the effective action of Eq.~\eqref{effective_action} is the absence of the topological term $(N/2)\;{\rm str}\;T^{-1}\partial T$ which, according to Refs.~\cite{Altland15prb,Altland01nulear,Altland14prl}, would lead to the delocalized zero modes for odd $N$. In other words, vanishing topological term in Eq.~\eqref{effective_action} implies Anderson localization for both even and odd $N$. From the effective action in \eq{effective_action}, it is conceptually straightforward to calculate the physical observables. For instance, the conductance at a given energy $E$ is the functional average of the corresponding retarded and advanced Green functions $g(E)\equiv \avg{G(E^+)G(E^-)}$. However, the actual evaluation using the supersymmetric nonlinear $\sigma$ model is technically complex and we just show the result here. Using the transfer matrix method \cite{Altland01nulear}, we obtain the conductance $g$ at zero energy for $L\gg \tilde \xi$:
\begin{equation}\label{conductance}
g\approx\sqrt{\frac{\tilde\xi}{\pi L}}\exp\Big[-\frac{L}{\tilde\xi}\Big],
\end{equation}
which is consistent with the numerically observed Anderson localization behavior. Moreover, from \eq{conductance}, it is clear that the coupling constant $\tilde \xi$ in the effective action of  \eq{effective_action} can be identified as the localization length, which
scales linearly with $N$ for $N\gg 1$. This linear-$N$ localization length for $N\!\gg\! 1$ is consistent with the result obtained from numerical calculations.

\begin{figure}[tpb]
\centering
\includegraphics[width=0.5\columnwidth]{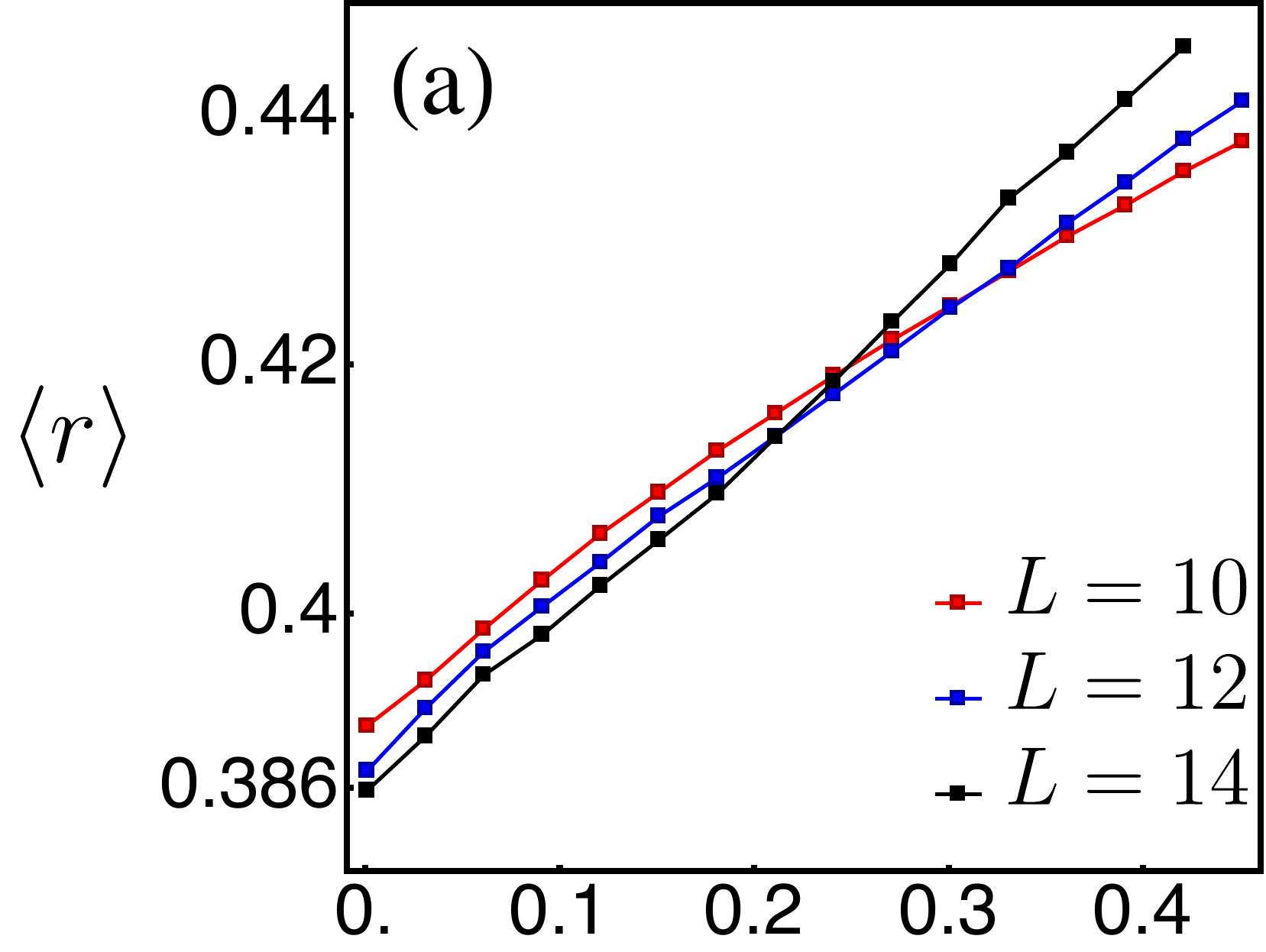}~~~
\includegraphics[width=0.44\columnwidth]{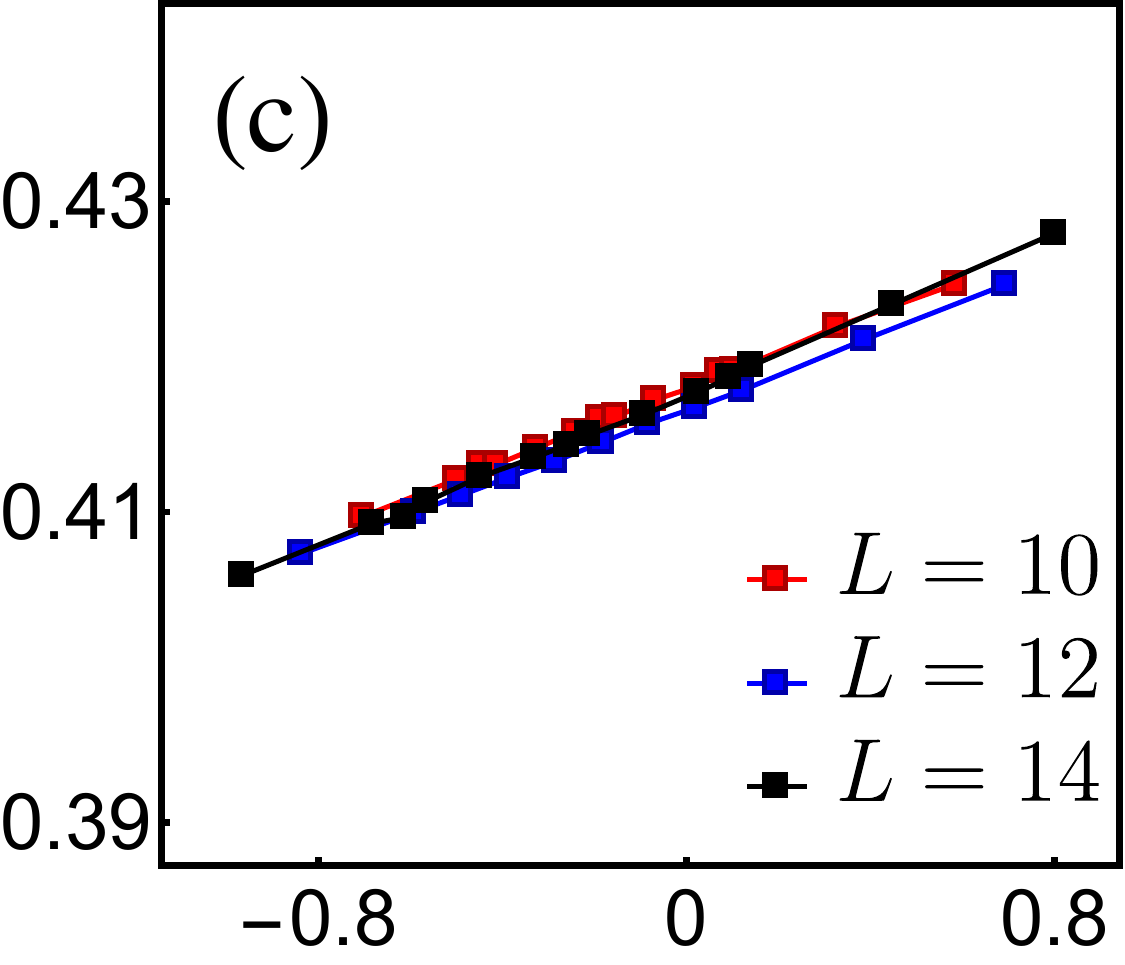}\\
\includegraphics[width=0.5\columnwidth]{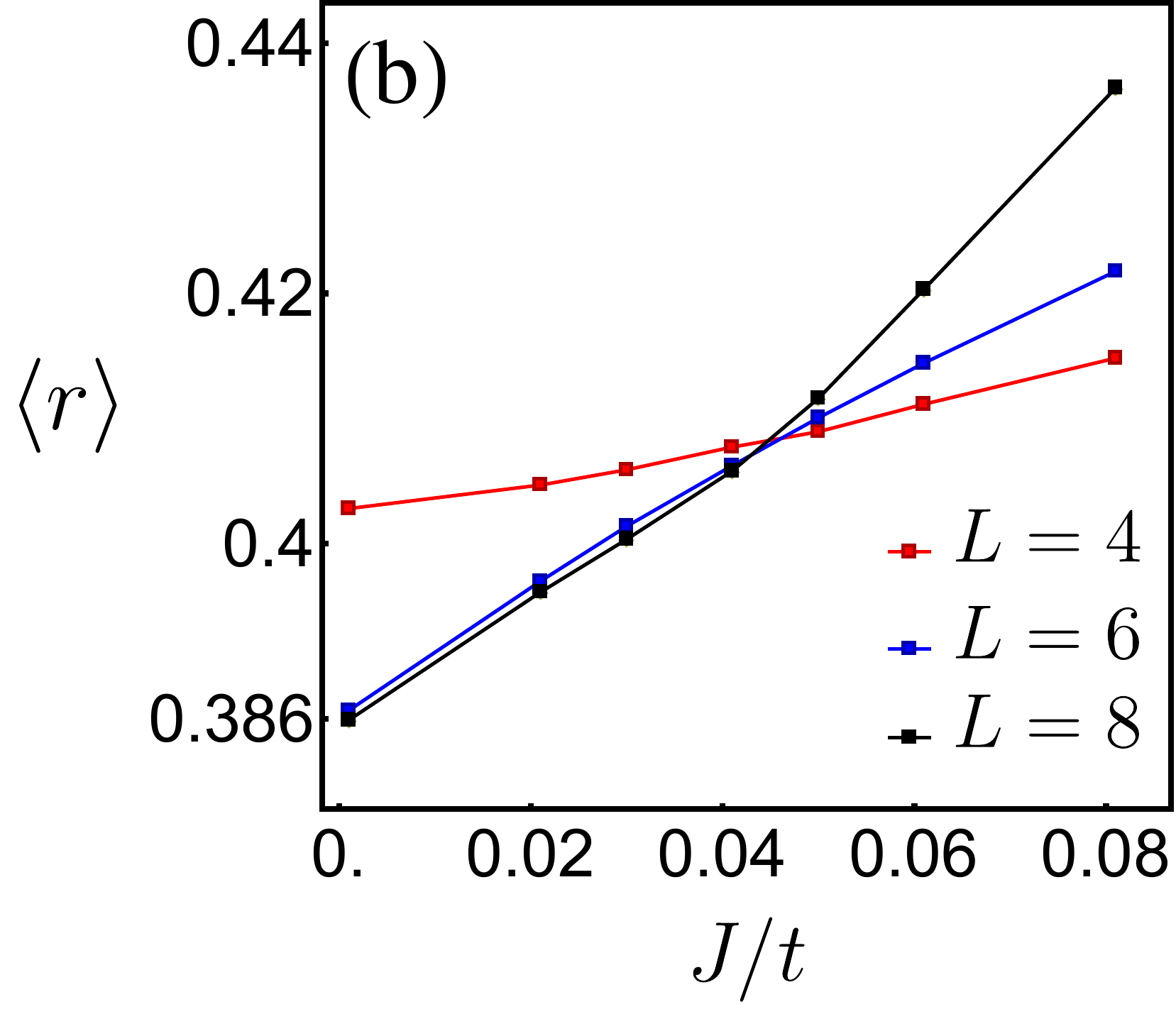}~~~
\includegraphics[width=0.44\columnwidth]{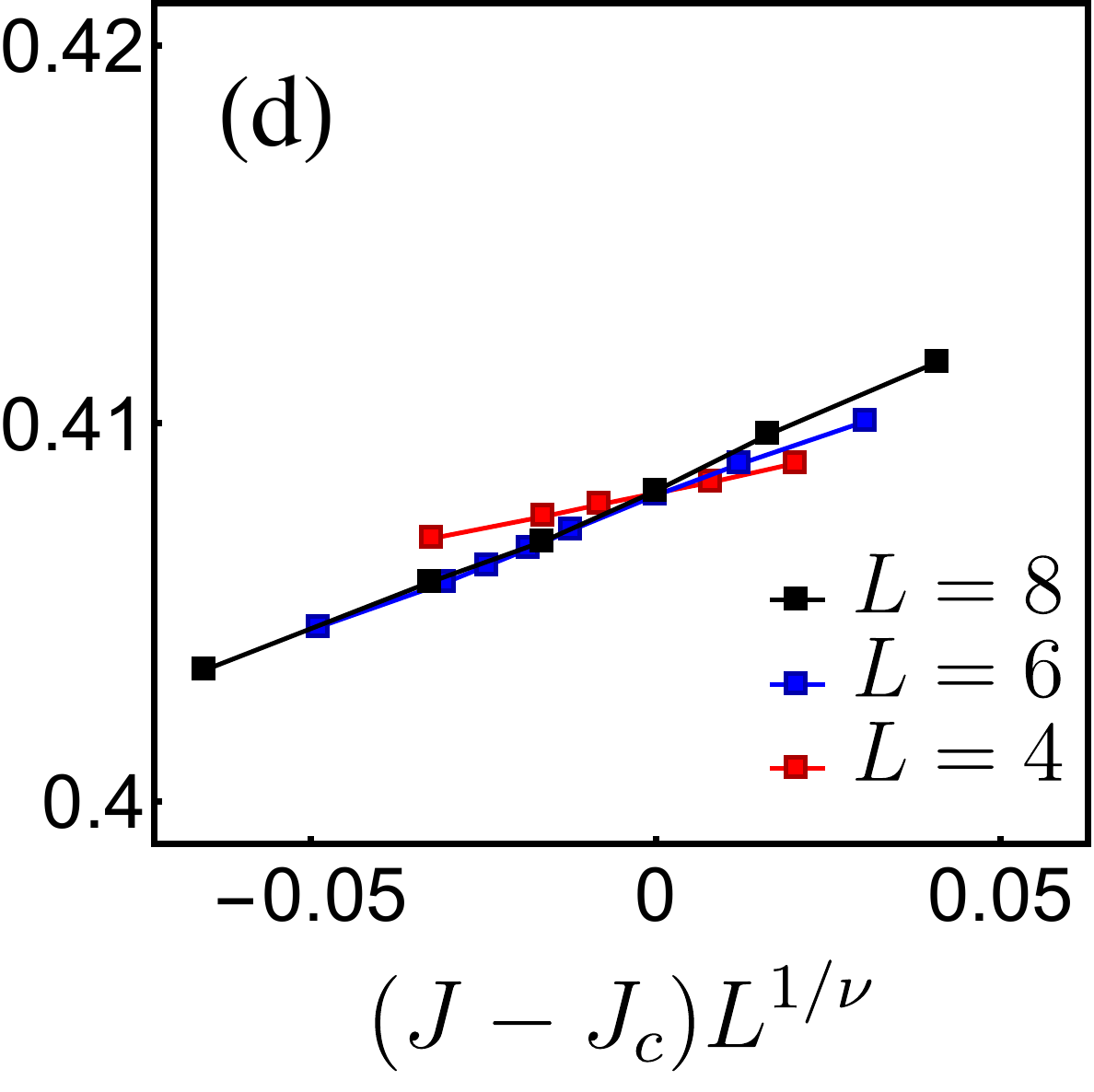}
\caption{Disorder averaged level statistics for $N=2$ (a) and $N=4$ (b).
By varying the interaction strength $J/t$, there are crossings between adjacent
system size $L$. The positions of the crossings gradually drift to
smaller $J/t$ but the trend of slowing down with
increasing $L$ can be seen clearly.
The finite-size data collapse for $N=2$ (c)
and $N=4$ (d) in the vicinity of the crossing points.
$J_c=0.23t, \nu=1.1$  for $N=2$, and $J_c=0.045t, \nu=0.99$ for $N=4$.
All the results are obtained by setting $t_1=0.5t,t_2=1.5t$.}
\label{level_statistics}
\end{figure}

{\bf Finite SYK interactions:} After establishing Anderson localization in the noninteracting limit, we are ready to consider finite interaction strength, i.e., $J\!>\!0$. To investigate how the interactions can thermalize the system, we employ ED to calculate the many-body level statistics of the interacting Hamiltonian of Majorana fermions in \eq{realSYK}. Assuming that $\{e_n\}$ denotes the many-body energy level in an ascending order, we calculate the dimensionless ratio $\widetilde{r}_n$ defined by $\widetilde r_n= \frac{\textrm{min}(s_n,s_{n-1})}{\textrm{max}(s_n,s_{n-1})}$, where $s_n=e_{n+1}-e_n$ \cite{huse07prb,Atas13prl}. For the uncorrelated energy levels obeying Poisson distribution, $\avg{\widetilde r}\to 2\ln 2-1\approx 0.386$; while for the Gaussian orthogonal ensemble (GOE) of random matrix, $\avg{\widetilde r}\to 0.53$. When $J\!=\!0$, $\avg{\widetilde r} \approx 0.386$ for $N\!=\!2$ and $4$, as shown in Fig.~\ref{level_statistics}(a) and (b), respectively, indicating Poisson distribution that is consistent with the Anderson localized state for finite $N$. Moreover, $\avg{\widetilde r}$ increases as $J$ increases, indicating that the SYK interactions tend to thermalize the system. For both $N\!=\!2$ and $4$, it is clear that $\avg{\tilde{r}}$ between adjacent system sizes $L$ crosses at a critical interaction strength $J_c$, indicating that there is a dynamic quantum phase transition from the MBL phase $(J\!<
\!J_c)$ to the thermalized phase ($J\!>\!J_c$). Due to the finite-size effect, the crossing points drift gradually towards smaller $J_c$ as $L$ increases, which is common in the ED studies of many-body localizations~\cite{huse07prb}. Nonetheless, the tendency of the drift becomes slower for larger $L$. Essentially, this implies that $J_c$ is nonzero and the MBL phase should persist below $J_c$ in the thermodynamic limit.

To characterize the MBL transitions, we explore the critical behaviors of the dynamic transition. Around the MBL transition, $\avg{\tilde r}$ should obey a universal scaling function, i.e., $\avg{\tilde r} = f[(J-J_c)L^{1/\nu}]$, where $\nu$ is the correlation/localization length critical exponent. By collapsing the data, as shown in Fig.~\ref{level_statistics}(c), we obtain the critical exponent $\nu \!\approx\! 1.1\pm 0.1 $ for $N\!=\!2$.
(for $N\!=\!4$, the data collapse shown in Fig. \ref{level_statistics}(d) gives rise to $\nu\!\approx\! 0.99\pm 0.2$). In order to improve the quality of data collapse, we set the variance of the random hopping in  Eq.~\eqref{realSYK} to be $t_1^2/N$ and $t_2^2/N$, $t_1 \!\ne\! t_2$, for odd and even bonds, respectively. The staggered variances significantly shorten the localization length (which is still proportional to $N$, as shown in the SM); accordingly, the finite-size effect decreases for the accessible system size. Our ED calculations show that the critical exponent $\nu \approx 1.1$, which is consistent with previous ED works on MBLwith previous ED works transition using small system size~\cite{Khemani17prl,Luitz15prb, Pollmann14prl} and is still quite different from the results obtained by real-space renormalization studies using large system size \cite{Vosk15prl,Potter15prx,Potter17prl,SXZhang18prl,SXZhang19arX}.

As explicitly shown for the $N\!=\!2$ and $4$ SYK chain, the finite-$N$ effect renders the MBL phase when the interaction strength $J$ is smaller than a critical value $J_c$. The value of $J_c$ of $N\!=\!4$ is smaller than the one of $N\!=\!2$, indicating that $J_c$ decreases as $N$ increases. Due to the absence of Anderson localization for $N\!=\!\infty$, it is clear that $J_c\!=\!0$ for $N\!=\!\infty$. As the discussion of the SYK models generally relies on a large-$N$ approximation to control the quantum fluctuations, it is interesting to further explore how the critical strength $J_c$ scale with $1/N$. In the weakly interacting limit, the energy scale corresponding to the MBL transition is given by $T_c \!\sim\! \frac{\delta_\xi}{\lambda |\log \lambda|}$ \cite{Gornyi05prl, basko06annal}, where $\delta_\xi \!=\! \frac1{\rho \xi}$ is the average level spacing of single-particle states within a localization length in the noninteracting limit. $\rho$ is the average density of single-particle states per unit volume, and the dimensionless quantity $\lambda\!=\! \frac{J}{N^{3/2}\delta_\xi}$ characterizes the interaction strength with respect to the average single-particle level spacing. It is known from the noninteracting calculations that $\xi \propto N$ and the average density of states per unit volume is found to be $\rho\!\propto\! t^{-1} N$ (see Supplemental Material), thus $\delta_\xi \!\propto\! t N^{-2}$ and $T_c\!\propto\! \frac{t^2}{J} \frac1{N^{5/2}\log N}$. It directly leads to a rough estimate of the critical interaction strength $J_c/t \!\propto\! \frac1{N^{5/2}\log N}$ for the dynamic transition of full many-body localization (namely requiring $T_c\sim t$ where $t$ is the order of the bandwidth). By using the numerical data shown in Fig.~\ref{level_statistics}, we estimate that the critical strength scales as $J_c \!\propto\! N^{-\eta}$ with $\eta\!\approx\! 2.4$, which is close to the scaling behavior of $\eta\!=\!5/2$ derived from the weakly interacting limit (up to a logarithmic correction). Note that this scaling is consistent with the requirement that $J_c$ vanishes at $N\!=\!\infty$.

{\bf Discussion and concluding remarks:} We have shown that, in the noninteracting limit, all the single particle states in the SYK chain at finite $N$ ($N$$\ge$$2$) are localized irrespective to the parity of $N$, due to the vanishing topological $\theta$-term. Here we conjecture that the same localization physics should apply to the other four symmetry classes in one dimension based on the notion of {\it superuniversality} \cite{Gruzberg05prb, Altland14prl, Altland15prb}, which refers to the fact that in one dimension, all five symmetry classes, including classes D and DIII, share similar low-energy properties. We further showed that the system enters an MBL phase for weak SYK interactions but undergoes a dynamic phase transition from the MBL phase to a thermalized phase when the interaction $J$ exceeds a critical value $J_c$ with $J_c/t\!\sim\! \frac{1}{N^{5/2}\log N}$. Finally, we mention some future directions related to finite $N$.  For instance, it would be desired to characterize the thermal phase at finite-$N$ in full details including its Lyapunov exponent, specific heat, and transport behaviors. Due to the finite-$N$ effect, it is expected that its characters should be renormalized from its large-$N$ limit.

{\it Acknowledgment.}---We thank E. Altmann, Y.-F. Gu, S. A. Kivelson, and X.-L. Qi for helpful discussions. This work is supported in part by the National Natural Science Foundation of China under Grant No. 11825404 (X.D., S.-K.J., and H.Y.), the Ministry of Science and Technology of China under Grant No. 2016YFA0301001 and No. 2018YFA0305604 (H.Y.), the Strategic Priority Research Program of Chinese Academy of Sciences under Grant No. XDB28000000 (H.Y.), Beijing Municipal Science and Technology Commission under Grant No. Z181100004218001 (H.Y.), and Beijing Natural Science Foundation under Grant No. Z180010 (H.Y.).

\begin{widetext}
\section{The Supplemental Materials}
\renewcommand{\thefigure}{S\arabic{figure}}
\setcounter{figure}{0}

\renewcommand{\theequation}{S\arabic{equation}}
\setcounter{equation}{0}
\renewcommand{\thesection}{S\arabic{section}}

\renewcommand{\thetable}{S\arabic{table}}

\subsection{A. Derivation of supersymmetric field theory}
\subsubsection{1. Disorder average}
The derivation of the supersymmetric field theory
largely follows the approach
developed in Refs.~\cite{Altland99nuclearB,Altland01nulear,Altland15prb}.
The hopping matrix elements satisfies
\begin{eqnarray}
&&\langle t_{nm}^{\mu\nu} \rangle = 0,\\
&&\langle t_{nm}^{\mu\nu} t_{nm}^{\nu'\mu'}\rangle = \frac{\lambda^2}{N}\delta_{\mu\mu'}
 \delta_{\nu\nu'}\delta_{m,n+1}.
\end{eqnarray}
In order to carry out the disorder average,
we promote the fermionic field $\phi$ to the
2-component superfield
\begin{equation}
\psi=
\begin{pmatrix}
\psi_b\\
\psi_f
\end{pmatrix}
\end{equation}
with the subscripts $b,f$ denote
the bosonic and fermionic field variables, respectively.
Then we can proceed by integrating over $t$,
\begin{equation}\label{disorder_average}
\begin{split}
&\avg{\exp\left(i\sum_{n\in{A},m\in{B},\mu\nu}\bar\psi_{n,\mu}t_{nm}^{\mu\nu}\psi_{m,\nu}+h.c.\right)}
=\mathcal{C}\int dt\exp
\left(i\sum_{n\in{A},m\in{B},\mu\nu}\bar\psi_{n,\mu}t_{nm}^{\mu\nu}\psi_{m,\nu}+h.c.
-\frac{N}{2 \lambda^2}{\rm Tr}\;t^2\right)\\
&=\mathcal{C}\int dt\exp\left(-\sum_{n\in{A},m\in{B},\mu\nu}
\abs{\sqrt{\frac{N}{2}}\frac{1}{\lambda}t_{nm}^{\mu\nu}
-i \lambda\sqrt{\frac{2}{N}}\bar\psi_{n,\mu}\psi_{m,\nu}}^2\right)
=\exp\left(-\sum_{n\in{A},m\in{B},\mu\nu}\frac{2\lambda
^2}{N}
\bar\psi_{n,\mu}\psi_{m,\nu}\bar\psi_{m,\nu}\psi_{n,\mu}\right)\\
&=\exp\left(-\sum_{n\in{A},m\in{B},\mu\nu}\frac{2\lambda^2}{N}
\psi_{n,\mu}\bar\psi_{n,\mu}\psi_{m,\nu}\bar\psi_{m,\nu}\right)
=\exp\left(-\frac{2\lambda^2}{N}\sum_{n\in{A},m\in{B},\mu\nu}
{\rm str\;} g^{\mu\mu}_n g^{\nu\nu}_m\right)
\end{split}
\end{equation}
with $\mathcal{C}$ being a normalization constant.
And we have introduced
the bilinear term
\begin{equation}
g^{\mu\mu}_n\equiv \psi_{n,\mu}\otimes\bar\psi_{n,\mu}
\end{equation}
In the last two identities of Eq.~\eqref{disorder_average}
we have made use of the cyclic invariance
property of the supertrace~\cite{wegner2016book}.
Then we arrive at the partition function Eq.~(3) in the main text.

\subsubsection{2. Hubbard-Stratonovich transformation}
Now we perform the Hubbard-Stratonovich transformation
by introducing a pair of supermatrix fields $Q^{\pm}_{nm}\equiv Q_{A,n}\pm iQ_{B,m}$,
with $Q_{A,n}$($Q_{B,m}$) lives on $A$($B$) sublattice, respectively.
\begin{equation}
\begin{split}
Z &= \int {\cal D}(\bar\psi,\psi)
\exp\left(i\sum_{n,\mu} \bar \psi_{n,\mu}z\psi_{n,\mu}
-\frac{2\lambda^2}{N}\sum_{n\in{A},m\in{B},\mu\nu}
{\rm str\;} g^{\mu\mu}_n g^{\nu\nu}_m\right)\\
&\times\int {\cal D}Q^{\pm}
\exp\left(-\sum_{n\in{A},m\in{B},\mu\nu}
\left(\frac{1}{\lambda}\sqrt{\frac{1}{2N}}Q^--i \lambda\sqrt{\frac{2}{N}}
\psi_{n,\mu}\bar\psi_{n,\mu}\right)
\left(\frac{1}{\lambda}\sqrt{\frac{1}{2N}}Q^+-i \lambda\sqrt{\frac{2}{N}}
\psi_{m,\nu}\bar\psi_{m,\nu}\right)\right)\\
&=\int {\cal D}Q^{\pm}{\cal D}(\bar\psi,\psi)
\exp\left(
i\sum_{n,\mu}\bar\psi_{n,\mu} z\psi_{n,\mu}+\frac{i}{N}\sum_{n\in{A},\mu\nu}
\bar\psi_{n,\mu}(Q^+_{n,n-1}+Q^+_{n,n+1})\psi_{n,\mu}\right.\\
&\left.+\frac{i}{N}\sum_{m\in{B},\mu\nu}\bar\psi_{m,\nu}(Q^-_{m,m-1}+Q^-_{m,m+1})\psi_{m,\nu}
-\frac{N}{2\lambda^2}\sum_{n\in A,m\in B}Q^+_{nm} Q^-_{mn}\right)
\end{split}
\end{equation}
The next step is to integrate out $\psi$
and we arrive at
\begin{eqnarray}\label{action}
S[Q^{\pm}] &=& \frac{N}{2t^2}
\sum_n {\rm str} (Q^+ Q^-)-N\sum_{n\in A}{\rm str}\ln (z+Q^+_{n,n+1}+Q^+_{n,n-1})
-N\sum_{m\in B}{\rm str}\ln (z+Q^-_{m,m+1}+Q^-_{m,m-1}).
\end{eqnarray}

\subsubsection{3. The non-linear \texorpdfstring{$\sigma$}{Lg}-model in the strongly disordered limit}
It is clear that, for $z=0$, the action in Eq.~\eqref{action} is invariant under the transformation $Q^+\to T_1 Q^+ T_2$ and $Q^-\to T_2^{-1} Q^- T_1^{-1}$, where $T_1,T_2\in \GL$, and $\GL$ is the generalization of the original fermionic symmetry. The overall factor $N$ enables us to seek the saddle point solution which is exact in the large-$N$ limit.
By assuming a uniform ansatz $Q^{\pm}=\frac{1}{2}(Q^{\pm}_{n,n+1}+Q^{\pm}_{n,n-1})$, from the saddle point condition ($\frac{\delta S}{\delta Q^{\pm}}=0$) we obtain
\begin{equation}
Q^{\mp}=\frac{2\lambda^2}{z+Q^{\mp}}\Longrightarrow Q^{\pm}_{\rm sp}
=\frac{1}{2}\left(-z\pm\sqrt{z^2+8\lambda^2}\right).
\end{equation}
To identify the low energy degrees of
freedom for $z=0$, we can parameterize $Q^{\pm}$ by $(Q^+,Q^-)=(PT,T^{-1}P)$
in Eq.~\eqref{action},
where both $T,P\in \GL$ and $T$ stands for massless
fluctuation while $P$ is the massive fluctuation
that is incompatible with the symmetry of the ground-state.

Let's ignore the massive fluctuations by setting $P=\openone$,
the action is of the form,
\begin{equation}\label{massless}
\begin{split}
 S_{\rm fl}[T]&=N\sum_{n\in A} {\rm str \, ln \,}
\left(T_{n,n+1}+T_{n,n-1}\right)+N\sum_{m\in B} {\rm
  str \, ln \,}
 \left(T^{-1}_{m,m+1}+T^{-1}_{m,m-1}\right).
\end{split}
\end{equation}
we then expand $T_{nm}$ as
\begin{equation}
  \label{taylor}
  T_{nm} = T_n + \frac{a}{2} \partial_{n,m} T_n +
  \frac{a^2}{8}\partial^2_{n,m} T_n +\dots
\end{equation}
where $a$ is the lattice constant and
$\partial_{n,m}$ denote the directional derivative
from site $n\to m$.
Taking Eq.~\eqref{taylor} into Eq.~\eqref{massless},
\begin{equation}\label{gradient}
\begin{split}
\frac{1}{N}S_{\rm fl}[T]&\approx\sum_{n\in A} {\rm str \, ln \,}
\left(2T_n+ \frac{a}{2}\partial_{n,n+1}T_n
+\frac{a}{2}\partial_{n,n-1}T_n
+ \frac{a^2}{8}\partial_{n,n+1}^2 T_n
+\frac{a^2}{8}\partial_{n,n-1}^2 T_n\right)\\
+&\sum_{m\in B} {\rm str \, ln \,}
 \left(2T_m^{-1}+ \frac{a}{2}\partial_{m,m+1}T_m^{-1}
 +\frac{a}{2}\partial_{m,m-1}T_m^{-1}
+ \frac{a^2}{8}\partial_{m,m+1}^2 T_m^{-1}
+\frac{a^2}{8}\partial_{m,m-1}^2 T_m^{-1}\right)\\
\approx& \sum_{n\in A} {\rm str \, ln \,} 2T_n
-\sum_{m\in B} {\rm str \, ln \,} 2T_m
+ \frac{a^2}{16}\sum_{n\in A}\left(T_n^{-1}\partial_{n,n+1}^2 T_n
+T_n^{-1}\partial_{n,n-1}^2 T_n\right)
+\frac{a^2}{16}\sum_{m\in B}\left(T_m\partial_{m,m+1}^2 T_m^{-1}
+T_m\partial_{m,m-1}^2 T_m^{-1}\right)\\
\approx& \frac{a^2}{16}\sum_{n\in A}\left(T_n^{-1}\partial_{n,n+1}^2 T_n
+T_n^{-1}\partial_{n,n-1}^2 T_n\right)
+\frac{a^2}{16}\sum_{m\in B}
\left(T_m\partial_{m,m+1}^2 T_m^{-1}+T_m\partial_{m,m-1}^2 T_m^{-1}\right),
\end{split}
\end{equation}
where we have made use of the fact
that $\sum_{m\in B} \partial_{n,m}T_n=0$.
By taking the continuum limit
$\sum_{n\in A} \to \frac{1}{2a} \int$,
Eq.~\eqref{gradient} can be written as
\begin{eqnarray}
S_{\rm fl}[T] &=& \frac{N a^2}{8}\left(\sum_{n\in A}
{\rm str }(T_{n}^{-1}
\partial^2 T_n)  +
\sum_{m\in B} {\rm str }(T_{m}
\partial^2 T_m^{-1})\right)\nonumber \\
&\simeq&\frac{N a}{16}\int {\rm str }\left(T^{-1}\partial^2 T +
    T\partial^2 T^{-1}\right)=-\frac{N a}{8}\int {\rm str}( \partial T^{-1} \partial T ).
\label{Sfl}
\end{eqnarray}
where the integration by parts is used in the last equality.

\subsection{B. The level statistics at large \texorpdfstring{$J/t$}{Lg} and onsite \texorpdfstring{$U$}{Lg}}
As shown in Fig.~\ref{level_statistics_large}, as $J/t$ increases, the $\avg{r}$ value increases
towards the GOE value 0.531, for both $N=2$ and $N=4$ .
\begin{figure}[b]
\centering
\includegraphics[width=0.31\columnwidth]{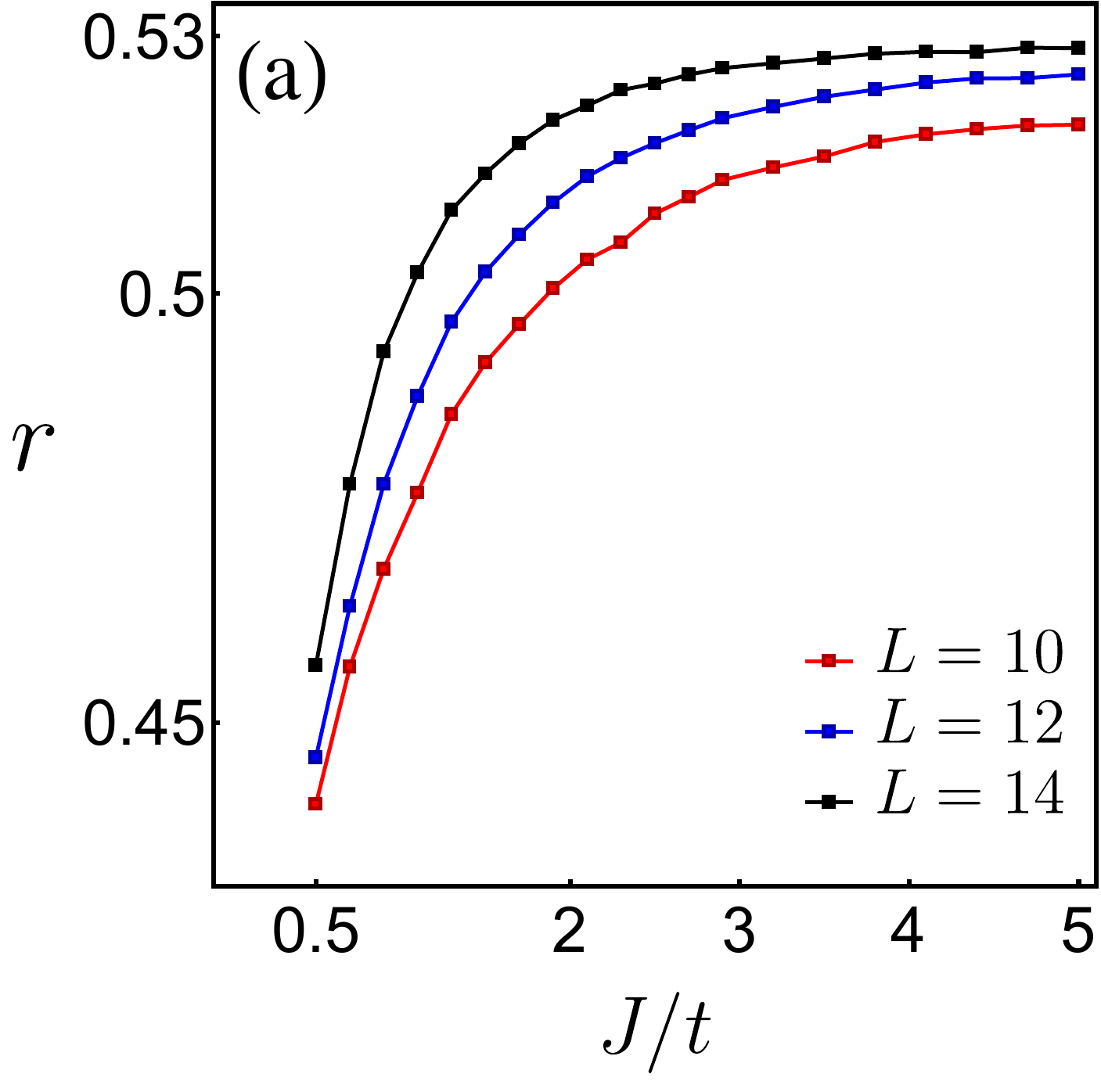}~~~~
\includegraphics[width=0.29\columnwidth]{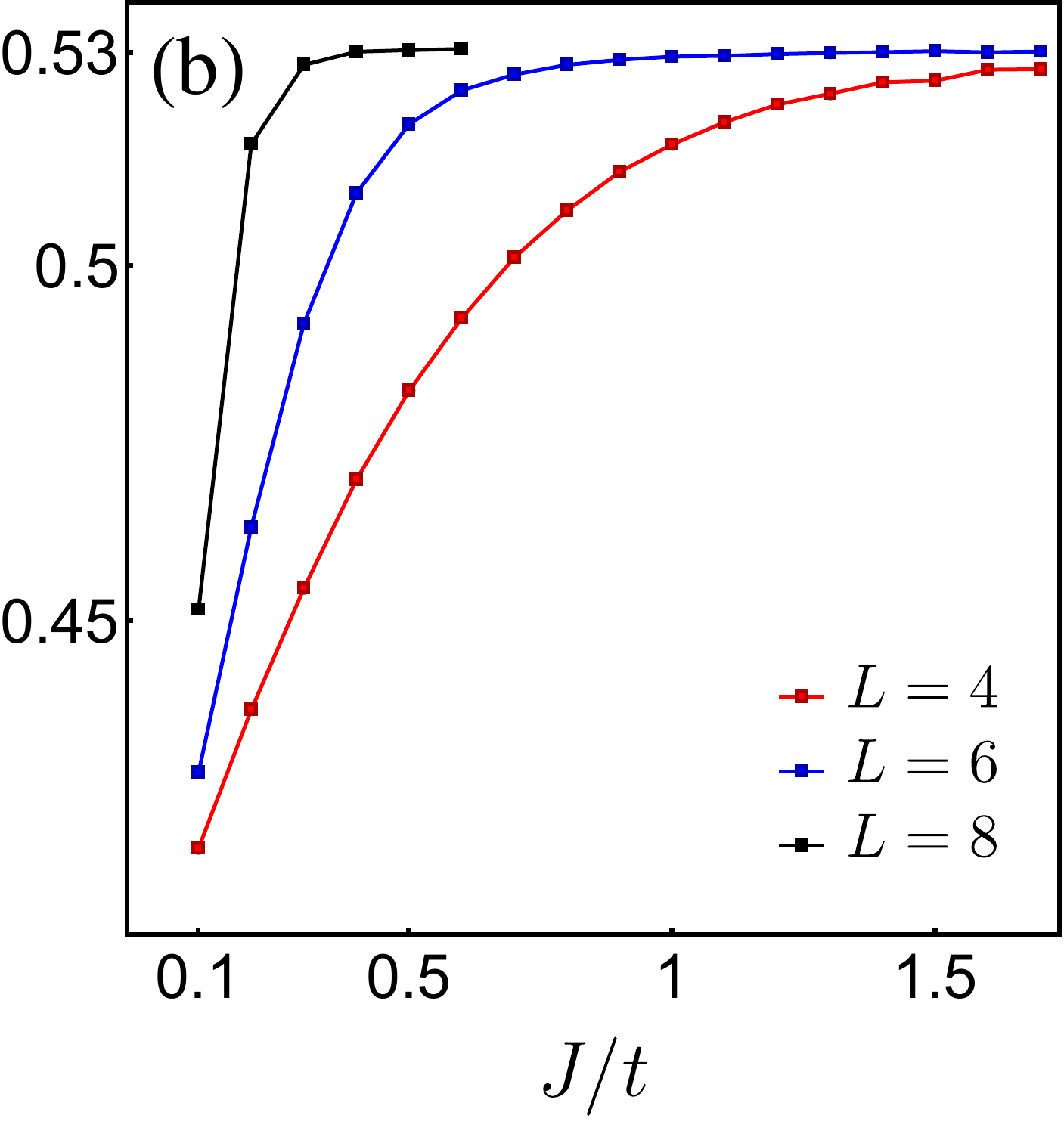}
\caption{The level statistics at large $J/t$ for $N=2$ (a) and $N=4$ (b).}
\label{level_statistics_large}
\end{figure}

Now we instead consider the onsite SYK interaction $U$ while setting $J=0$. As shown in Fig.~\ref{level_statistics_U}(a), the $\avg{r}$ value asymptotes to Possion and GOE value at small and large $U/t$
limit, respectively. If we zoom in, as shown in Fig.~\ref{level_statistics_U}(b), we can see there is a crossing around $U/t\approx 0.13$, indicating a dynamical MBL-ETH phase transition.

\begin{figure}[htb]
\centering
\includegraphics[width=0.3\textwidth]{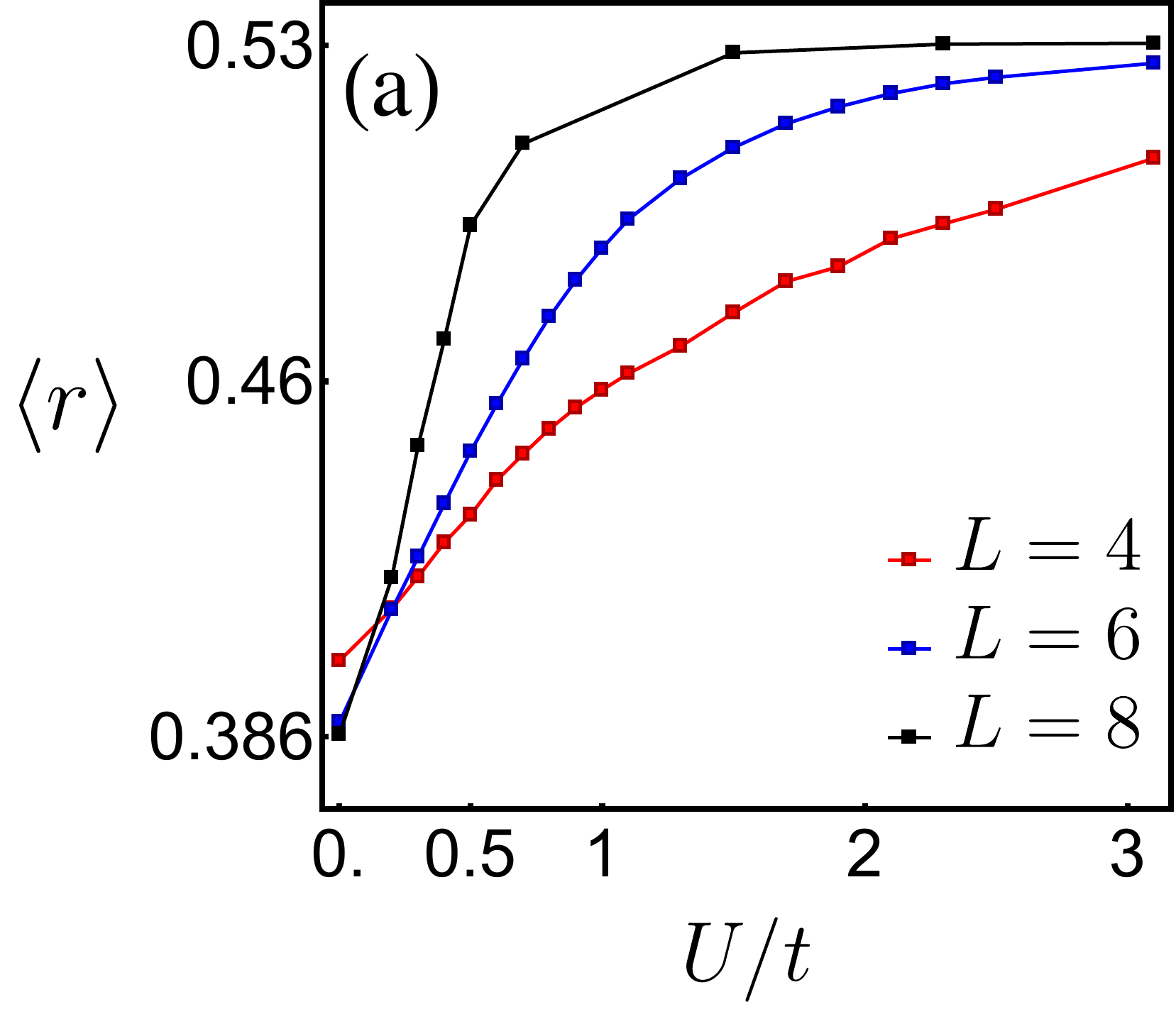}~~~~
\includegraphics[width=0.26\textwidth]{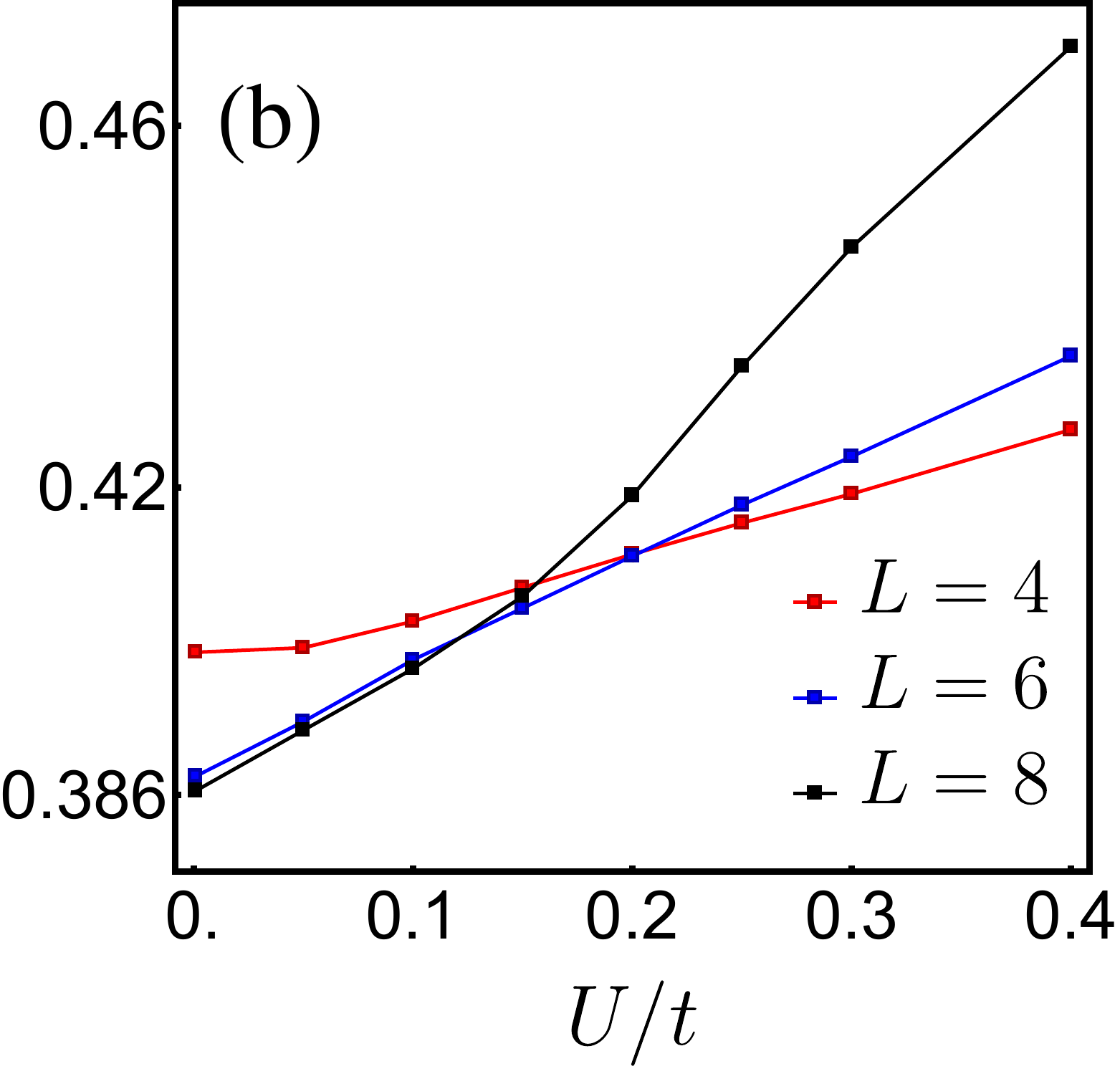}
\caption{The level statistics as a function of $U/t$ for $N=4$ and $J=0$. }
\label{level_statistics_U}
\end{figure}

\subsection{C. Density of States and localization length in the noninteracting limit}
\begin{figure}[htb]
\centering
\includegraphics[width=0.30\columnwidth]{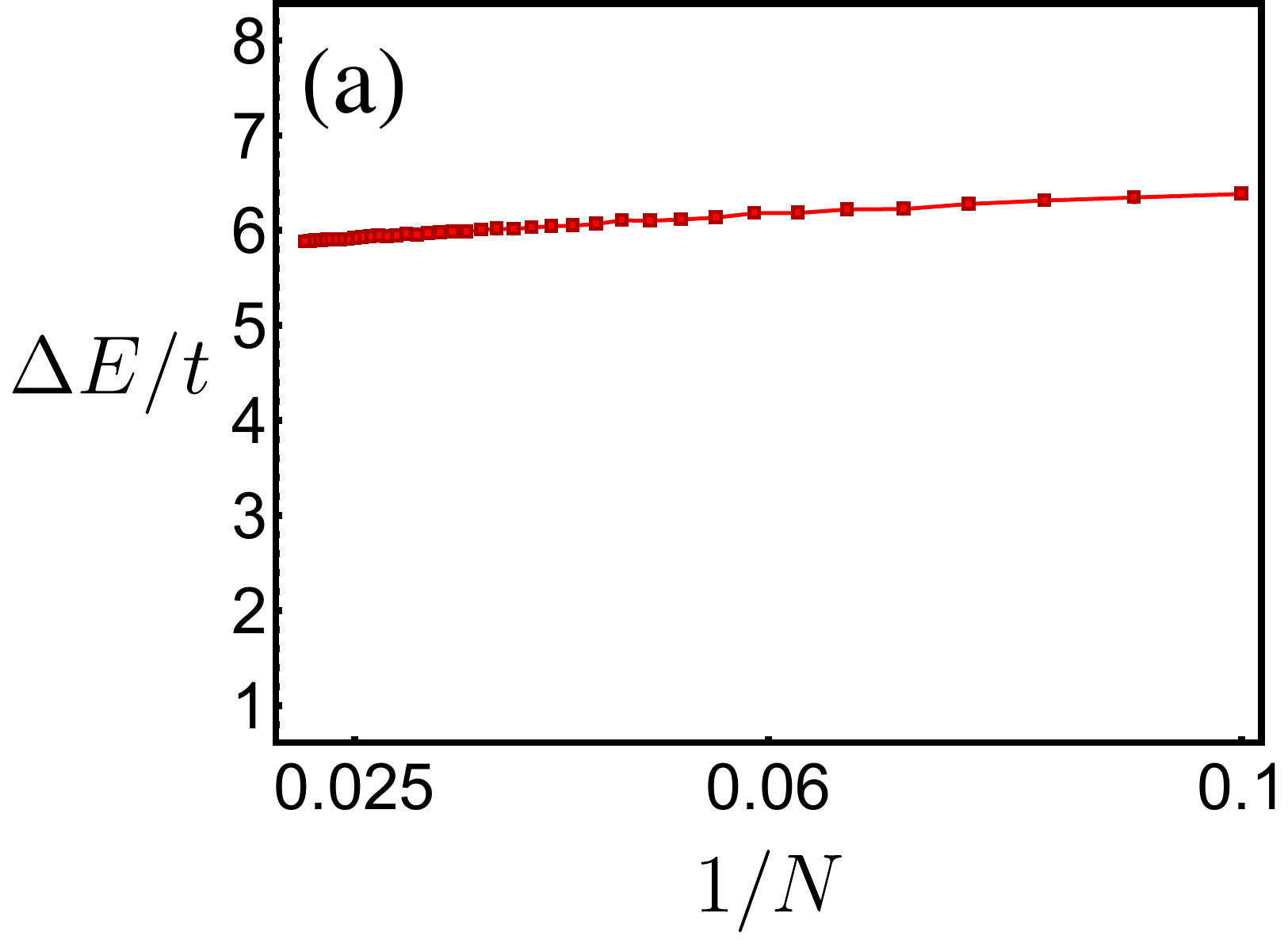}~~~~
\includegraphics[width=0.26\columnwidth]{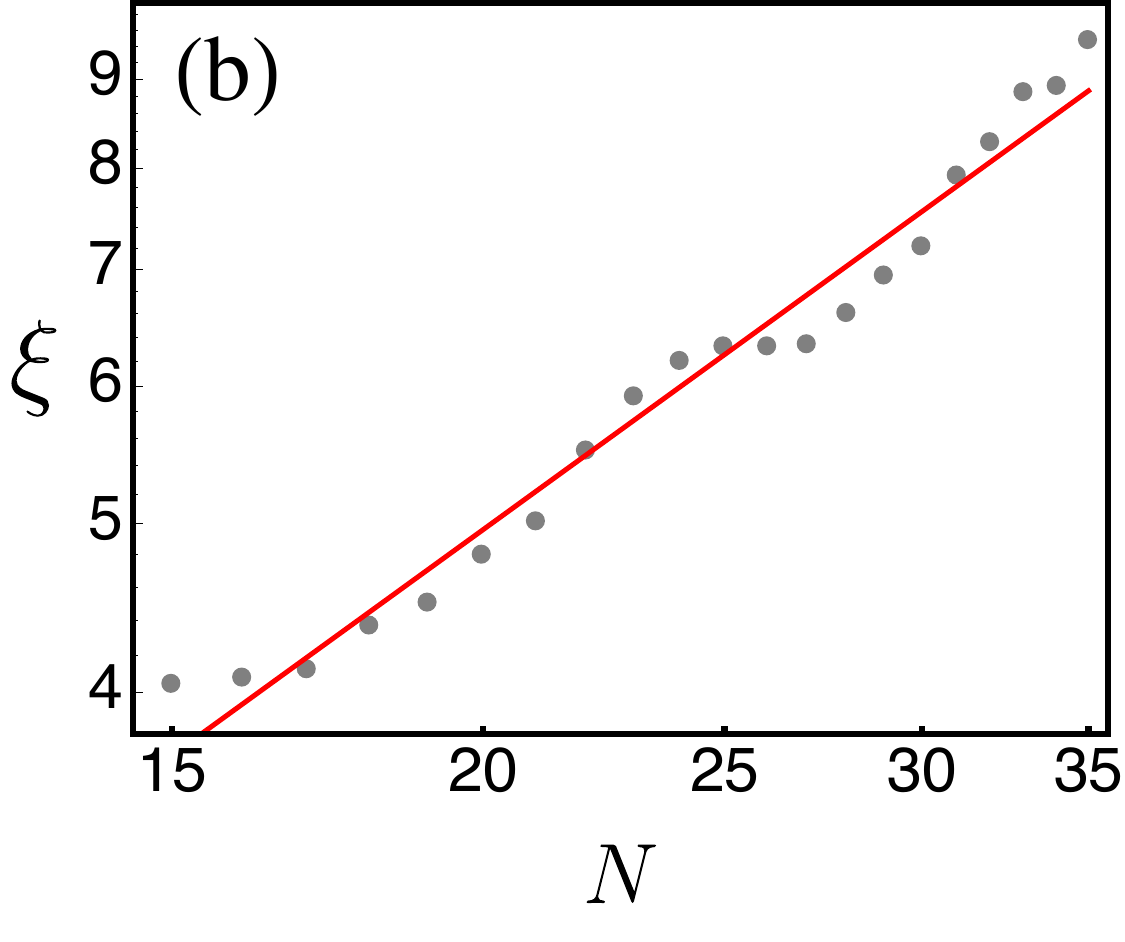}
\caption{(a) $\Delta E$ as a function of $1/N$ with $N\in[10,48]$. (b)The log-log plot for the localization length $\xi$ as a function of $N$ with $N\in[15,35]$. The system size $L=1001$. All the results are obtained by setting $t_1=0.5t$ and $t_2=1.5t$.}
\label{dos}
\end{figure}
In the noninteracting limit, there are $N\times L$ single particle
states in total.
Therefore the single particle density of states $\rho$ per unit length
can be found as
\begin{equation}
\rho=\frac{N L}{L}\frac{1}{\Delta E}
=\frac{N}{\Delta E},
\end{equation}
where $\Delta E$ is the total bandwidth. As shown in Fig.~\ref{dos}(a), $\Delta E/t$ saturates to constant as $N\to \infty$ with fixed $L$. Consequently, we conclude that $\rho\propto N/t$.

In addition, we also computed the localization length $\xi$ in the presence of dimerization. The data shown in Fig.~\ref{dos}(b) gives rise to $\xi\approx 0.22N^{\alpha}$ with $\alpha=1.04\pm 0.04$. While in the uniform case mentioned in the main text we have $\xi\approx 0.38N^{1.02\pm0.02}$. So we find in both cases $\xi$ always scales linearly with $N$ and the dimerization effectively shortens the localization length $\xi$.

\end{widetext}


%

\end{document}